# Atomic Layer Deposition of Metal Oxides on Pristine and Functionalized Graphene

Xinran Wang, Scott Tabakman and Hongjie Dai[*]

*Department of Chemistry and Laboratory for Advanced Materials, Stanford University, Stanford, CA 94305, USA*
RECEIVED DATE (automatically inserted by publisher); Email: hdai@stanford.edu

Graphene has recently emerged as an interesting material for electronics due to extremely high carrier mobility in bulk graphene[1] and the demonstration of all-semiconducting sub-10nm graphene nanoribbons[2]. Aggressive device scaling requires integration of ultrathin high-κ dielectrics in order to achieve high on-state current and ideal subthreshold swing without substantial gate leakage[3]. Deposition of metal oxides including high-κ dielectrics on graphene has not been systematically investigated thus far. Uniform oxide deposition on graphene by atomic layer deposition (ALD) is expected to be difficult due to the lack of dangling bonds in the graphene plane, as in the case of carbon nanotube[4,5]. As a result, previously reported topgated graphene devices used very thick (≥15nm) ALD dielectrics on top of negative tone resist[6] or functional layer[7] to prevent gate leakage. Functionalization is likely needed for uniform ALD on graphene.

Here we show that ALD of metal oxides gives no direct deposition on defect-free pristine graphene. On the edges and defect sites however, dangling bonds or functional groups can react with ALD precursors to afford active oxide growth. This leads to an interesting and simple way to decorate and visualize defects in graphene. By non-covalent functionalization of graphene with carboxylate terminated perylene molecules, one can coat graphene with densely packed polar groups for uniform ALD of high-κ dielectrics. Uniform high-κ coverage is achieved on large pieces of graphene sheets with size of greater than 5μm. This method opens the possibility of integrating ultrathin high-κ dielectrics in future graphene electronics.

Our graphene sheets were obtained by standard peel-off method on 300nm $SiO_2$/Si substrate described in Ref. 8. We first identified few-layer (≤5 layers) graphene sheets under an optical microscope. Then the chip was annealed at 600ºC in vacuum with 1Torr argon atmosphere to clean the substrate and graphene sheets, followed by atomic force microscopy (AFM) imaging of the few layer graphene sheets. Then the chip was soaked in 3,4,9,10 perylene tetracarboxylic acid (PTCA) solution for ~30mins, thoroughly rinsed and blown dry (see Supporting Information). The chip was immediately moved into the ALD chamber to prevent contamination by molecules in the air. We then deposited $Al_2O_3$ at ~100 ºC using trimethylaluminum and water as precursors[9]. In the same run we also included a $SiO_2$/Si chip covered by PMMA except some pre-patterned squares to measure $Al_2O_3$ thickness (See Supporting Information).

We first carried out study on pristine graphene. Figure 1a and b show AFM images of the same area before and after ALD of ~2nm $Al_2O_3$. Before ALD, the height of the triangular graphene piece at the bottom and the large piece on the left was ~1.7nm and ~2.0nm, respectively. Near the edge of the 2.0nm graphene, there was also a narrow ~1.0nm high stripe. After ALD, ~2.0nm $Al_2O_3$ was coated on $SiO_2$, the apparent topography height of the three graphene sheets was obviously reduced, to a similar level as the ALD coated $SiO_2$. The height difference before and after ALD indicates no $Al_2O_3$ coating on pristine graphene sheets. This is because ALD relies on chemisorption or rapid reaction of precursor molecules with surface functional groups[5,10]. Since pristine graphene does not have any dangling bonds or surface groups to react with precursors, no ALD occurs on the graphene plane. Interestingly, we observed quasi continuous bright lines of $Al_2O_3$ preferentially grown on the edges of graphene sheets, suggesting dangling bonds on the edges or possible termination by –OH or other reactive species (Figure 3b). We also observed some bright dots in the middle of graphene sheets, likely corresponding to defects such as pentagon-hexagon pairs or vacancies known to exist in graphite[11] (Figure 3c). Thus, our ALD method could be used as a novel way to probe and visualize defects in graphene, which is much simpler and more efficient than current STM[12] and TEM[11] measurements. The peeled off graphene sheets could be defect free for a few micrometers as evidenced by AFM images after ALD (see Supporting Information). However, defects do exist in the graphene planes and should be considered in other studies.

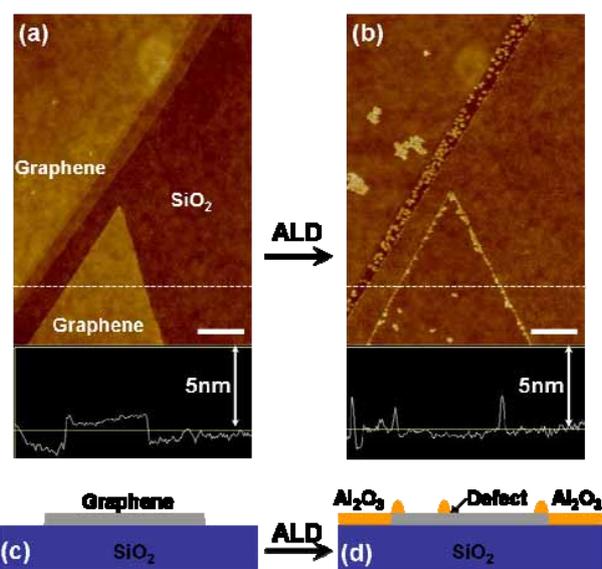

*Figure 1.* ALD of $Al_2O_3$ on pristine graphene. (a) AFM image of graphene on $SiO_2$ before ALD. The height of the triangular shaped graphene is ~1.7nm as shown in the height profile along the dashed line cut. Scale bar is 200nm. (b) AFM image of the same area as (a) after ~2nm $Al_2O_3$ ALD deposition. The height of the triangular shaped graphene becomes ~ 0.3nm as shown in the height profile along the dashed line cut. Scale bar is 200nm. (c) and (d) are schematics of graphene on $SiO_2$ before and after ALD. The $Al_2O_3$ grows preferentially on graphene edge and defect sites.

In order to afford uniform ALD coating on pristine graphene, functionalization of graphene is needed to induce uniform surface groups as active ALD nucleation sites. PTCA is an excellent candidate for selective coating of graphitic surfaces on $SiO_2$/Si substrate owing to its planar, conjugated ring system and its symmetrically arranged, negatively-charged terminal carboxylates (Figure 3). While PTCA non-covalently partitions and adheres to graphene, likely via π-π stacking and hydrophobic forces[14,15], basic conditions deprotonate the terminal acid groups, yielding a tetra-anionic state which is repulsed from the oxidized silica surface. Following adsorption onto graphitic surfaces, the carboxylate functional groups on the perylene moiety serve as uniformly distributed sites for nucleation of ALD (Figure 3d).

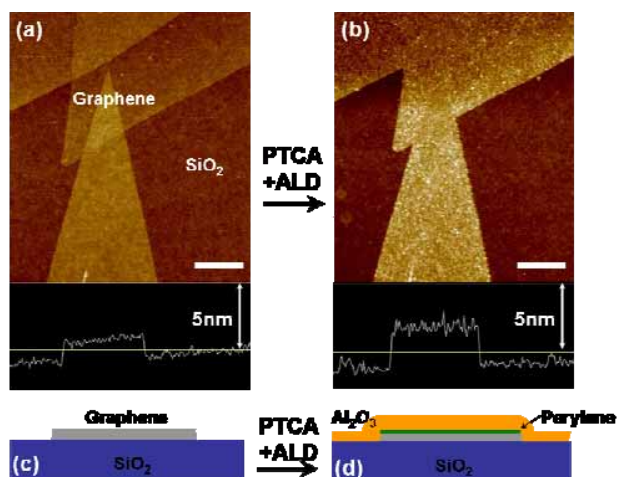

*Figure 2.* ALD of $Al_2O_3$ on PTCA coated graphene. (a) AFM image of graphene on $SiO_2$ before ALD. The height of the triangular shaped graphene is ~1.6nm as shown in the height profile along the dashed line cut. Scale bar is 500nm. (b) AFM image of the same area as (a) after ~2nm $Al_2O_3$ ALD deposition. The height of the triangular shaped graphene becomes ~3.0nm as shown in the height profile along the dashed line cut. Scale bar is 500nm. (c) and (d) are schematics of graphene on $SiO_2$ before and after ALD. The $Al_2O_3$ grows uniformly on non-covalently PTCA coated graphene.

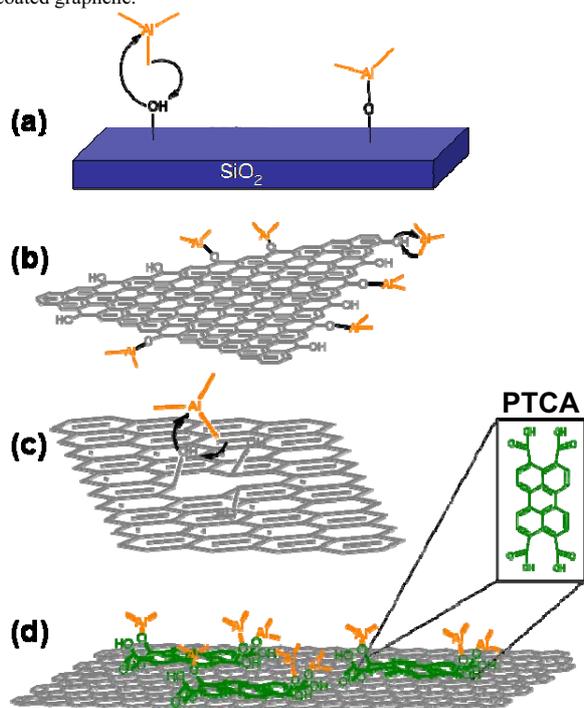

*Figure 3.* Schematics of atomic layer deposition of aluminum oxide via trimethylaluminum (TMA) precursor on (a) bare $SiO_2$ substrate, (b) a single-layer graphene sheet with edges, (c) graphene containing a defect site, and (d) perylene-tetracarboxylic acid (PTCA) coated graphene. PTCA selectively adheres to graphene on SiO2 surfaces, providing binding sites for TMA deposition. Inset is a top view of PTCA structure.

Figure 2a and b show AFM images of the same area for PTCA treated sample before and after ~2nm $Al_2O_3$ deposition. Before ALD, the graphene under the line cut was ~1.6nm high, while after ALD, the height increased to ~3.0nm as shown in the height profile in Figure 2b. The relative height increase of $Al_2O_3$ coated graphene was partly attributed to the thickness of PTCA layer, which was usually ~0.5-0.8nm as observed by AFM after PTCA coating step (see Supporting Information). The actual $Al_2O_3$ on graphene was ~2.8±0.2nm thick in Figure 2b. Complete and uniform PTCA packing is evidenced by $Al_2O_3$ coated over the whole piece of graphene, which is several micrometers in size. The mean roughness of $Al_2O_3$ film on graphene is ~0.33nm over 2.5μm x 2.5μm area as measured with several pieces (see Supporting Information). Previous high vacuum STM studies have confirmed two modes of epitaxial packing of the PTCA precursor, perylene-tetracarboxic dianhydride, along the lattice lines of highly oriented pyrolytic graphite[14,15]. As the partitioning of PTCA from methanol to the graphene interface should be highly favorable, it is likely that similar epitaxial packing is occurring in the solution phase, yielding dense, uniform coating of graphene sheets. This self-assembly process of PTCA on graphene is responsible for uniform $Al_2O_3$ film coating. Note that non-covalent functionalization for ALD is not expected to degrade the electrical properties of graphene, similar to the case of single-walled carbon nanotubes[4].

In summary, we found that ALD of metal oxide cannot directly be deposited on pristine graphene without surface functionalization due to the lack of dangling bonds and surface functional groups. ALD could grow actively on edge and defect sites of graphene, which could be used as a simple and effective probe to defects. We used carboxylate terminated perylene molecules to functionalize graphene with densely packed –OH groups and achieved uniform ultrathin $Al_2O_3$ deposition on graphene over a few micrometers. The non-covalent functionalization method is not destructive to graphene and could be used for depositing ultrathin high-κ dielectrics for future graphene electronics.

**Acknowledgement.** This work was supported by Intel and Marco MSD Focus Center.

**Supporting Information Available:** ALD process details, measurement of ALD height using pre-patterned PMMA windows, $Al_2O_3$ film surface roughness, and process of making PTCA solution are described in online supporting information. This material is available free of charge via Internet at http://pubs.acs.org.

# Atomic Layer Deposition of Metal Oxides on Pristine and Functionalized Graphene

Xinran Wang, Scott Tabakman and Hongjie Dai[*]

*Department of Chemistry and Laboratory for Advanced Materials, Stanford University, Stanford, CA 94305, USA*

## Supporting Information

**ALD process and measurement of $Al_2O_3$ thickness.** We used low temperature ALD process described in Ref. 9 in main text. The deposition was carried out at 100ºC in ~300mTorr pure nitrogen environment using trimethylaluminum (TMA) and water vapor as precursors. For each cycle of ALD, water pulse was 0.5s in duration, followed by a 40s purging time, a 0.5s TMA pulse and a 15s purging time. We always put a $SiO_2$/Si chip with pre-patterned 5μm x 5μm PMMA windows together with the real chips to measure the thickness of ALD deposition. After ALD, we did standard lift and measured the height of PMMA windows. The deposition rate of $Al_2O_3$ on $SiO_2$ was ~0.19nm/cycle in our ALD. Figure S1 is a typical AFM image after 10 cycles of ALD.

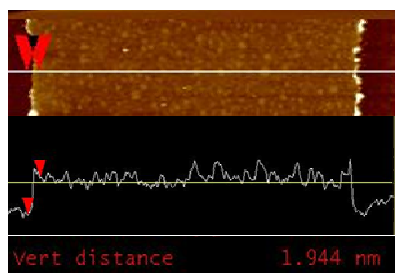

Figure S1. Measurement of $Al_2O_3$ thickness. The height was ~1.9nm after 10 cycles of ALD deposition.

**ALD deposition on pristine graphene.** We found that our peeled off graphene has relatively low density of defects far away from edges. Defect free area could be as large as a few micrometers. Figure S2 shows a pristine graphene piece before and after ALD, no defect was observed in ~3μm.

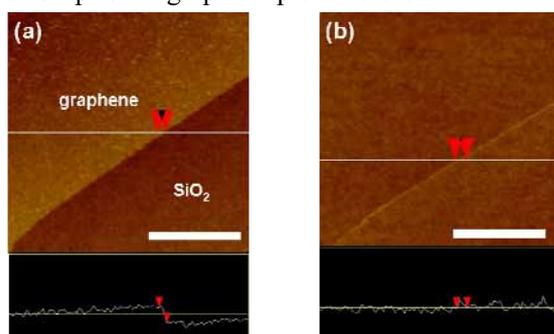

Figure S2. ALD on pristine graphene. AFM images of before (a) and after (b) ALD coating in the same area. No bright dots were observed on graphene except the edge. The graphene height is ~2.0nm and ~0nm before and after ALD, respectively. Scale bars are 1μm.

**Surface roughness of $Al_2O_3$ on PTCA coated graphene.** The roughness of $Al_2O_3$ is very good over a few micrometers. Figure S3 shows AFM images of $Al_2O_3$ on PTCA coated graphene surface. Mean roughness is ~0.33nm over the whole area.

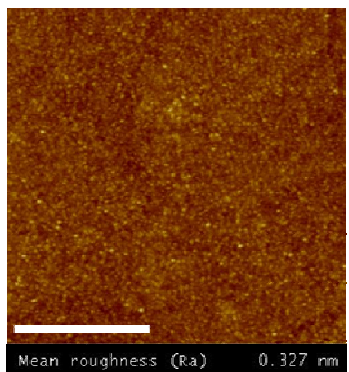

Figure S3. AFM image on a piece of PTCA coated graphene after ~2nm $Al_2O_3$ ALD. Very niform coating over 2.5μm x 2.5μm area was observed. The mean surface roughness is ~0.33nm. Scale bar is 1μm.

**Process of making PTCA solution.** PTCA solution was made by hydrolyzing 3,4,9,10 perylene-tetracarboxylic dianhydride (Fluka) in a minimal volume of 1M sodium hydroxide (Aldrich), thus yielding the tetracarboxylate. The yellow-green solution was then diluted in 7:1 methanol: 1M NaOH(aq) by volume, to a final concentration of 20μM perylene-tetracarboxylate and ~0.2M NaOH. The PTCA solution was poured over the graphene/$SiO_2$ chip and incubated for 30 minutes with shaking at room temperature. Finally, the substrate was cleaned by methanol and water rinses, and blown dry. Careful and thorough rinsing was needed to avoid multiple layer PTCA molecule stacking and uneven oxide deposition.

**Height increase after PTCA coating on graphene.** We observed ~0.5-0.8nm height increase after the PTCA treatment described in main text. This indicated that probably only a monolayer of closed packed PTCA is absorbed on graphene.

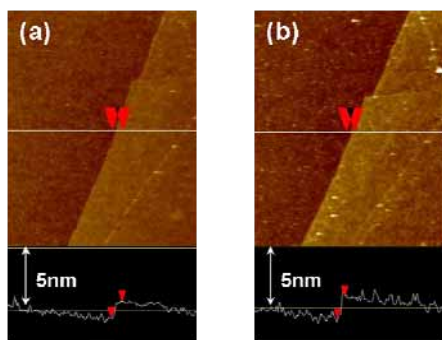

Figure S4. AFM images of the same graphene before (a) and after (b) PTCA coating. The height is ~1.4nm and ~1.9nm before and after PTCA coating, respectively.